# Universality of citation distributions revisited

Ludo Waltman, Nees Jan van Eck, and Anthony F.J. van Raan

Centre for Science and Technology Studies, Leiden University, The Netherlands
{waltmanlr, ecknjpvan, vanraan}@cwts.leidenuniv.nl

Radicchi, Fortunato, and Castellano [Radicchi, F., Fortunato, S., & Castellano, C. (2008). Universality of citation distributions: Toward an objective measure of scientific impact. *Proceedings of the National Academy of Sciences*, *105*(45), 17268–17272] claim that, apart from a scaling factor, all fields of science are characterized by the same citation distribution. We present a large-scale validation study of this universality-of-citation-distributions claim. Our analysis shows that claiming citation distributions to be universal for all fields of science is not warranted. Although many fields indeed seem to have fairly similar citation distributions, there are quite some exceptions as well. We also briefly discuss the consequences of our findings for the measurement of scientific impact using citation-based bibliometric indicators.

## 1. Introduction

In this paper, we present a validation study of earlier work by Radicchi, Fortunato, and Castellano (2008; see also Castellano & Radicchi, 2009; Radicchi & Castellano, 2011) on the universality of citation distributions. The number of citations of a publication can be rescaled by dividing it by the average number of citations of all publications that appeared in the same field of science and in the same year. Radicchi et al. (henceforth RFC) claim that the distribution of the rescaled citation scores of all publications in a certain field and a certain year is the same for all fields and all years. They refer to this phenomenon as the *universality of citation distributions*. According to RFC, the universality of citation distributions "justifies the use of relative indicators to compare in a fair manner the impact of articles across different disciplines and years" (p. 17271). Hence, the universality of citation distributions would provide a justification for the use of bibliometric indicators such as those discussed by Lundberg (2007) and Waltman, Van Eck, Van Leeuwen, Visser, and Van Raan (2011).

RFC's claim that citation distributions are universal is based on an analysis of 14 fields of science, where a field is defined by a journal subject category in the Web of Science database. In a follow-up paper on their initial work, Castellano and Radicchi (2009) emphasize "the need to validate the hypothesis of universality for all scientific disciplines (and not only a subset of them)" (p. 90). The aim of the present paper is to study the validity of the universality claim for all fields of science. We note that an earlier validation study of RFC's work was presented by Bornmann and Daniel (2009). However, this was a very limited study, since it was based on a rather small number of publications (i.e., fewer than 2000 publications, all in the field of chemistry). The validity of the universality claim is also investigated in a recent paper on the skewness of citation distributions (Albarrán, Crespo, Ortuño, & Ruiz-Castillo, 2011a, 2011b). This paper uses a different methodology than we do, but it arrives at a similar conclusion.

## 2. Data

We use data from the Web of Science (WoS) database for our analysis. We only consider publications that are classified as 'article' in WoS. Hence, publications classified as 'editorial material', 'letter', 'review', etc. are not taken into account. We collect data for 221 fields of science. Each field corresponds with a journal subject category in WoS. The 221 fields cover both the sciences and the social sciences. The arts and humanities are excluded from our analysis. We note that journal subject categories are overlapping. Some publications therefore belong to more than one field. These publications occur multiple times in our analysis, once for each field to which they belong.

For each publication, we count the number of citations received during the first ten years after the publication appeared (i.e., we use a ten-year citation window). For instance, in the case of a publication that appeared in 1999, citations are counted until the end of 2008.[1] We calculate the rescaled citation score of a publication as the number of citations of the publication divided by the average number of citations of all publications that appeared in the same field and in the same year. In the notation used by RFC, this is denoted by $c_f = c / c_0$, where $c$ denotes the number of citations of a publication, $c_0$ denotes the average number of citations of all publications in the same field and in the same year, and $c_f$ denotes the rescaled citation score.

Our analysis differs from the analysis of RFC in the following ways:
1. We study many more fields than RFC. Also, unlike RFC, we do not restrict ourselves to the sciences. We also consider the social sciences.
2. Unlike RFC, we do not exclude uncited publications from our analysis. We see no good reason for excluding these publications, and it is not clear to us why RFC have chosen to exclude them.
3. Unlike RFC, we do not include publications classified as 'letter' in our analysis. We prefer to leave out letters because their citation characteristics may be quite different from the citation characteristics of ordinary articles.

## 3. Results

Our analysis is based on publications that appeared in 1999. (RFC also mainly used publications from 1999 in their analysis.) We identified about 750,000 publications in WoS, each of them belonging to one or more fields. Out of the 221 fields in the sciences and the social sciences, there were 37 with fewer than 1,000 publications in 1999. These fields were excluded from our analysis. The remaining 184 fields have an average number of publications of 6,314. We found that the 90th percentile of the distribution of rescaled citation scores for all fields taken together equals 2.36.[2] Hence, 10% of all publications have a rescaled citation score that exceeds 2.36. We refer to these publications as *top 10% publications*. Our analysis focuses mostly on top 10% publications.

Clearly, if citation distributions are indeed universal, in each field approximately one out of ten publications should be a top 10% publication. Figure 1 shows a histogram of the distribution of the proportion top 10% publications for the 184 fields in our analysis. The figure also shows the theoretically expected distribution derived under the assumption that citation distributions are universal. More specifically, to derive the theoretically expected distribution, we assume that in a field with *n* publications the number of top 10% publications follows a binomial distribution with number of trials equal to *n* and success probability equal to 0.1. The theoretically expected distribution is obtained by aggregating the binomial distributions of the 184 fields. We note that RFC used the same theoretically expected distribution in their analysis (see RFC, p. 17269; see also Albarrán et al., 2011a, p. 18–19).

---

[1] In our analysis, the length of the citation window is the same for all fields. This is similar to the analysis of RFC. An alternative approach would be to adjust the length of the citation window to the citation characteristics of a field (cf. Stringer, Sales-Pardo, & Amaral, 2008).

[2] This value is somewhat higher than the value reported by Castellano and Radicchi (2009, Table 2). Assuming the distribution of rescaled citation scores to be lognormal, Castellano and Radicchi derived that the 90th percentile of the distribution equals 2.25. When we excluded uncited publications from our calculations (like Castellano and Radicchi did), we indeed found a value of 2.25.

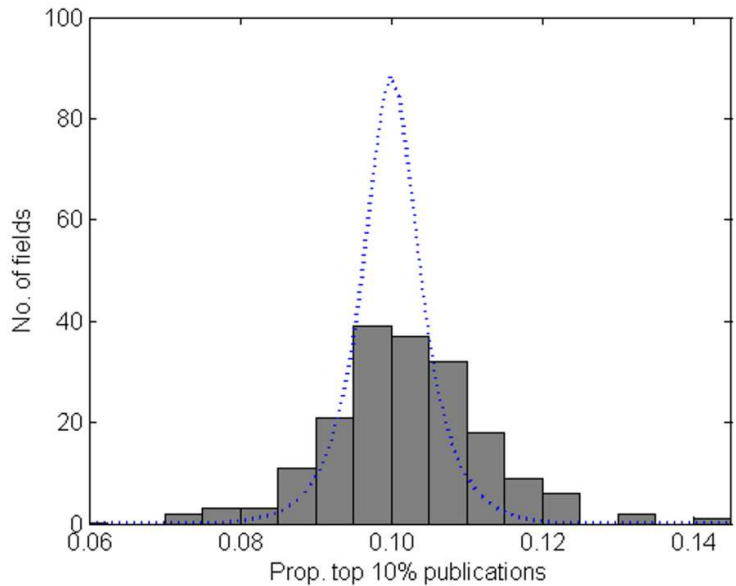

Figure 1. Histogram of the distribution of the proportion top 10% publications for 184 fields. The dotted curve indicates the theoretically expected distribution derived under the assumption that citation distributions are universal.

Figure 1 provides only limited support for the universality-of-citation-distributions claim. For instance, there turn out to be 55 fields in which the proportion top 10% publications is lower than 0.09 or higher than 0.11. Hence, in almost one-third of all fields, the proportion top 10% publications deviates more than 10% from the expected value of 0.10. According to the theoretically expected distribution, there should only be about 13 fields with a proportion top 10% publications below 0.09 or above 0.11. Furthermore, looking at the tails of the histogram in Figure 1, it can be seen that in some fields the proportion top 10% publications is more than 50% higher than in other fields.

Deviations from the universality of citation distributions can also be assessed by comparing the standard deviation of the proportion top 10% publications over the 184 fields with the theoretically expected standard deviation (i.e., the standard deviation of the theoretically expected distribution). In line with Figure 1, the empirically observed standard deviation turns out to be almost twice as high as the theoretically expected standard deviation (0.0105 vs. 0.0054). Table 1 shows that similar observations can be made when looking at top 5%, top 20%, and top 40% publications rather than at top 10% publications.

Table 1. Standard deviation of the proportion top 5%, top 10%, top 20%, and top 40% publications over the 184 fields. Both the empirically observed and the theoretically expected standard deviation are reported.

|  | St. dev. prop. top pub. ($\times 10^{-2}$) | Theoretically expected st. dev. ($\times 10^{-2}$) |
|---|---|---|
| Top 5% pub. | 0.99 | 0.39 |
| Top 10% pub. | 1.05 | 0.54 |
| Top 20% pub. | 1.74 | 0.71 |
| Top 40% pub. | 5.02 | 0.87 |

Table 2. Fields with the lowest proportion top 10% publications.

| Field | No. of pub. | Average no. of cit. | Prop. top 10% pub. |
|---|---|---|---|
| Crystallography | 5,620 | 9.56 | 0.070 |
| Mathematical & computational biology | 1,416 | 18.03 | 0.072 |
| Behavioral sciences | 2,858 | 20.04 | 0.075 |
| Biochemical research methods | 7,573 | 19.03 | 0.078 |
| Evolutionary biology | 2,564 | 28.86 | 0.079 |
| Marine & freshwater biology | 6,366 | 14.84 | 0.082 |
| Reproductive biology | 3,112 | 20.42 | 0.083 |
| Physics, atomic, molecular & chemical | 11,855 | 17.81 | 0.083 |
| Geography, physical | 1,832 | 15.47 | 0.086 |
| Geriatrics & gerontology | 1,701 | 17.82 | 0.087 |

Table 3. Fields with the highest proportion top 10% publications.

| Field | No. of pub. | Average no. of cit. | Prop. top 10% pub. |
|---|---|---|---|
| Materials science, paper & wood | 1,589 | 3.68 | 0.140 |
| Engineering, petroleum | 2,207 | 1.65 | 0.135 |
| Engineering, aerospace | 3,471 | 1.99 | 0.133 |
| Materials science, characterization & testing | 1,369 | 2.60 | 0.124 |
| Engineering, civil | 3,837 | 6.33 | 0.124 |
| Social issues | 1,039 | 5.50 | 0.122 |
| Multidisciplinary sciences | 9,392 | 65.33 | 0.122 |
| Anthropology | 1,571 | 7.55 | 0.122 |
| Materials science, ceramics | 3,709 | 8.00 | 0.120 |
| Social work | 1,069 | 8.03 | 0.119 |

Tables 2 and 3 list the ten fields with the highest and the lowest proportion top 10% publications. For each field, the number of publications, the average number of citations per publication, and the proportion top 10% publications are reported. For comparison, we note that the average number of citations per publication for all fields taken together equals 16.50. Comparing the fields listed in the two tables, some clear differences can be observed. Fields with a low proportion top 10% publications can be found mainly in the life sciences and the natural sciences (see Table 2). Most of these fields have an average number of citations per publication that is relatively close to the average found for all fields together. Fields with a high proportion top 10% publications can be found in the engineering sciences, the materials sciences, and the social sciences (see Table 3).³ These fields have a low average number of citations per publication. Based on this last observation, it seems that fields with a low average number of citations per publication tend to have a more skewed distribution of rescaled citation scores. Hence, there seems to be a tendency for fields with a low average number of citations per publication to deviate from the universality of citation distributions. This tendency can also be observed in Figure 2, in which a scatter plot is shown of the relation between a field's proportion top 10% publications and a field's average number of citations per publication.

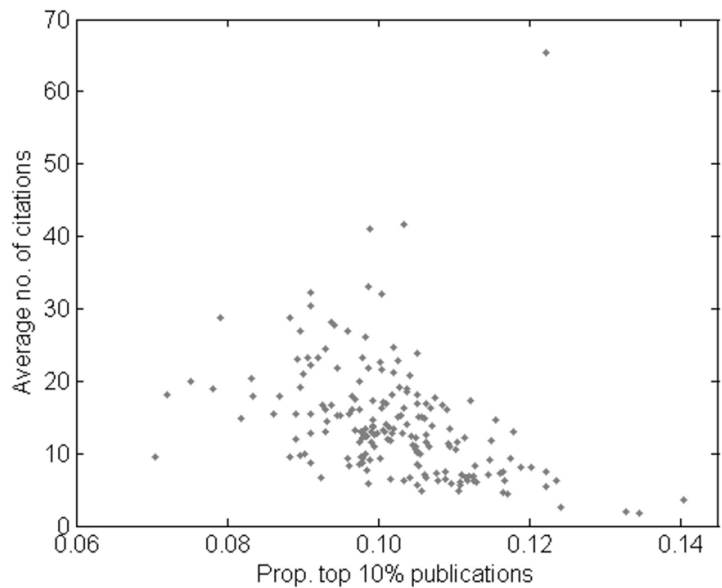

Figure 2. Scatter plot of the proportion top 10% publications vs. the average number of citations per publication for 184 fields ($\rho = -0.37$).

Figure 3 shows the cumulative distribution of rescaled citation scores for the three fields with the highest and the lowest proportion top 10% publications. The cumulative distribution for all fields taken together is shown as well. The figure clearly shows that different fields can have quite different citation distributions. This is partly due to differences in the proportion uncited publications. In some fields (e.g., *Engineering, petroleum* and *Engineering, aerospace*) about two-third of all publications are uncited, while in other fields (e.g., *Behavioral sciences*) there are almost no uncited publications. We will come back to the issue of the uncited publications later on.

---

³ *Multidisciplinary sciences* in Table 2 is an exception. The special characteristics of the *Multidisciplinary sciences* journal subject category were already noticed by RFC.

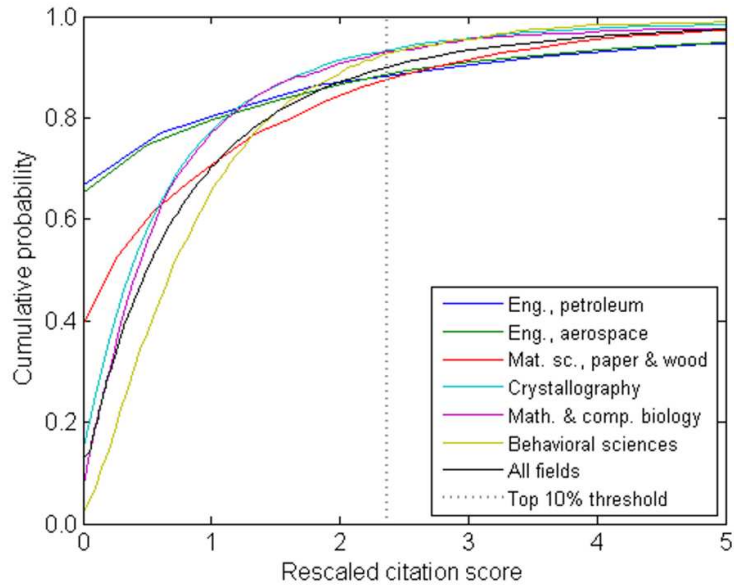

Figure 3. Cumulative distribution of rescaled citation scores for six selected fields and for all fields together.

To provide some additional evidence that deviations from the universality of citation distributions can only partly be explained by random effects, we repeated our analysis, but instead of publications that appeared in 1999 we used publications that appeared five years earlier, in 1994. Citations were again counted using a ten-year citation window. Figure 4 presents a scatter plot showing the relation between a field's proportion top 10% publications in 1999 and a field's proportion top 10% publications in 1994. As can be seen, the relation is fairly strong. Fields with a low (high) proportion top 10% publications in 1999 also tend to have a low (high) proportion top 10% publications in 1994.[4] Based on Figure 4, it can be concluded that random effects can only partly explain the observed deviations from the universality of citation distributions. For a considerable part, deviations are structural rather than random. This is in line with our earlier observation (based on Figure 1) that the distribution of the proportion top 10% publications for the 184 fields in our analysis is more dispersed than would be expected if citation distributions are universal.

---

[4] The same result was obtained when instead of publications from 1994 we used publications from 1995, 1996, 1997, or 1998.

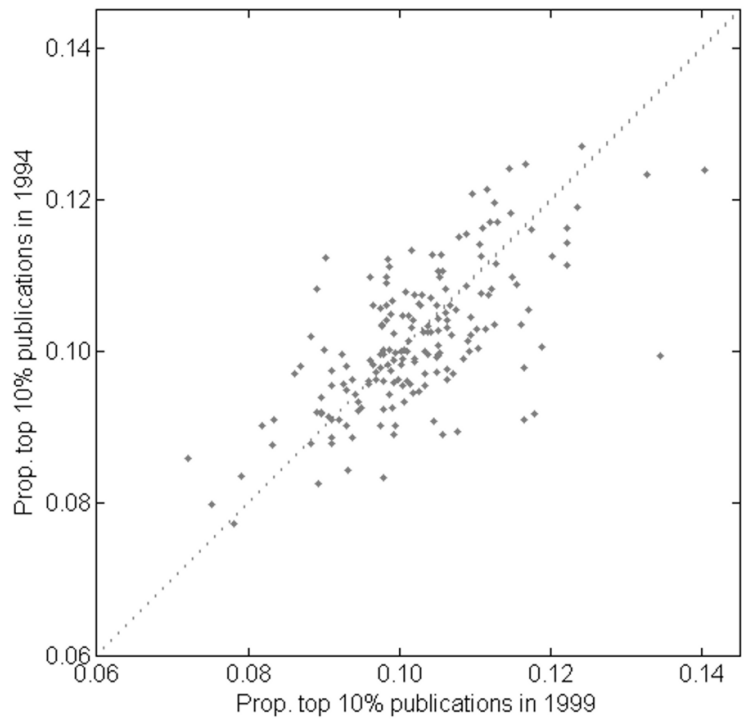

Figure 4. Scatter plot of the proportion top 10% publications in 1999 vs. the proportion top 10% publications in 1994 for 184 fields ($\rho = 0.69$).

Until now, we have used a ten-year time period to count the number of citations of a publication. This is similar to the time period that was used by RFC in most of their analysis. In practice, when citations are counted for research assessment purposes, one usually has to work with shorter time periods. This raises the question of the degree to which the validity of RFC's universality-of-citation-distributions claim depends on the length of the time period used for counting citations. To investigate this issue, we again repeated our analysis. Like in our original analysis, we used publications that appeared in 1999, but instead of counting citations during a ten-year time period, we counted citations during either a three-year or a five-year time period. For both time periods, we determined the distribution of the proportion top 10% publications for the 184 fields in the analysis. The distribution for a five-year time period (not shown) turns out to be slightly more dispersed than the distribution for a ten-year time period (shown in Figure 1), but the difference is small. Figure 5 shows a histogram of the distribution for a three-year time period. Comparing Figure 5 with Figure 1, it is clear that a three-year time period yields a much more dispersed distribution than a ten-year time period. When a three-year time period is used for counting citations, differences among fields in the proportion top 10% publications are rather large. The histogram in Figure 5 also deviates strongly from the theoretically expected distribution. Based on these observations, we conclude that the universality-of-citation-distributions claim has less validity when shorter time periods are used for counting citations.[5]

---

[5] We also performed an analysis in which citations were counted during a twenty-year time period. We used publications from 1989 for this analysis. The analysis yielded results similar to Figure 1.

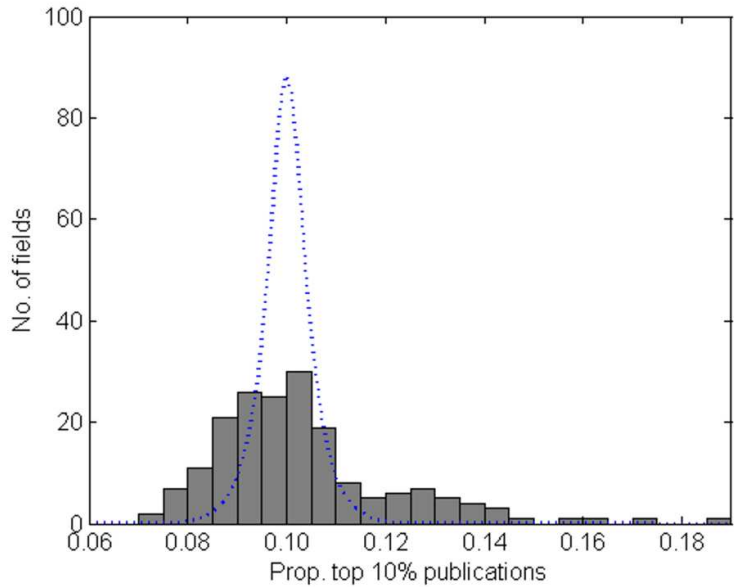

Figure 5. Histogram of the distribution of the proportion top 10% publications for 184 fields. Unlike in Figure 1, citations were counted during a three-year (rather than a ten-year) time period. The dotted curve indicates the theoretically expected distribution derived under the assumption that citation distributions are universal. Notice that the horizontal axis has a different scale than in Figure 1.

A question that still remains is how the differences between the results of our analysis and the results of the analysis of RFC can be explained. One part of the explanation is that our analysis is based on many more publications and fields than the analysis of RFC. Hence, our analysis is more comprehensive and therefore provides a more complete picture. Another part of the explanation (already suggested by Figure 3) may be that our analysis includes uncited publications while the analysis of RFC does not. To investigate the effect of excluding uncited publications, we also left out these publications from our own analysis. Without uncited publications, there were 176 fields with at least 1,000 publications in 1999. For each of these fields, we calculated the proportion top 10% publications. Figure 6 shows a histogram of the distribution of the proportion top 10% publications for the 176 fields. The distribution in Figure 6 is less dispersed than the distribution in Figure 1, as is also indicated by the standard deviations of the distributions (0.0074 and 0.0105, respectively). In Figure 1, 55 out of the 184 fields have a proportion top 10% publications that is lower than 0.09 or higher than 0.11. In Figure 6, this is the case for only 23 out of the 176 fields. We note, however, that this is still above the theoretical expectation. According to the theoretically expected distribution in Figure 6, there should be about 14 fields with a proportion top 10% publications below 0.09 or above 0.11. In line with this, the theoretically expected standard deviation is also somewhat lower than the empirically observed standard deviation (0.0056 vs. 0.0074). The comparison of the results obtained with and without uncited publications indicates that the effect of excluding uncited publications is quite substantial. When uncited publications are excluded, the claim that citation distributions are universal becomes more justifiable. We note that RFC made the following statement on the issue of the uncited publications: "Our calculations neglect uncited articles; we have verified, however, that their inclusion (…) does not affect the results of our analysis." (p. 17272). Our results make clear that this optimistic statement does not hold in general.

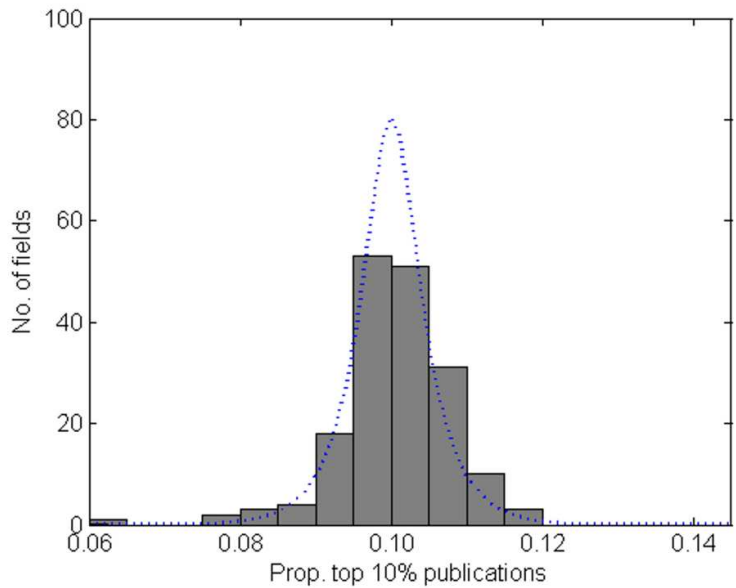

Figure 6. Histogram of the distribution of the proportion top 10% publications for 176 fields. Unlike in Figure 1, uncited publications were excluded from the analysis. The dotted curve indicates the theoretically expected distribution derived under the assumption that citation distributions are universal.

## 4. Conclusion

Our validation study provides only limited support for RFC's claim that citation distributions are universal. After appropriate rescaling, many fields of science indeed seem to have fairly similar citation distributions. However, there are quite some exceptions as well. Especially fields with a relatively low average number of citations per publication, as can be found in the engineering sciences, the materials sciences, and the social sciences, seem to have non-universal citation distributions. We found that deviations from the universality of citation distributions can only partly be explained by random effects. For a considerable part, deviations are structural rather than random. Based on the results of our analysis, we conclude that claiming citation distributions to be universal for all fields of science is not warranted. We note, however, that the universality claim becomes more justifiable when uncited publications are excluded from the analysis.

According to RFC, universality of citation distributions provides a justification for the use of relative bibliometric indicators. This raises the question whether the use of relative bibliometric indicators, such as those of Lundberg (2007) and Waltman et al. (2011), is still justified if citation distributions are not universal. In our opinion, the answer to this question is positive. We do not see universality of citation distributions as a necessary condition for the use of, for instance, our mean normalized citation score indicator (Waltman et al., 2011). Interpreting citation counts as approximate measures of scientific impact, non-universality of citation distributions may simply reflect that in some fields differences in the scientific impact of publications are larger than in other fields. For instance, some fields may be characterized by a small number of highly influential publications and a large number of much less influential publications, while other fields may have a more balanced distribution of scientific impact over publications. Correcting for such differences among fields, as proposed by some authors (Bornmann & Daniel, 2009; Lundberg, 2007), may be considered undesirable because it may distort the measurements of scientific impact provided by citation counts. In fields with a more dispersed distribution of scientific impact over publications, for instance, highly influential publications would not receive all the credits they deserve and too many credits would be given to less influential publications.

**Acknowledgment**

We would like to thank two referees for their comments on earlier drafts of this paper. The comments of the referees have led to a number of important improvements of the paper.